\documentclass[intlimits,twoside,a4paper]{article}

\usepackage[greek,english]{babel}
\RequirePackage{color}
\definecolor{MyDarkGreen}{rgb}{0.1,0.60,0.2}

\usepackage{amsmath,amssymb}
\usepackage{graphicx}

\usepackage[T2A]{fontenc}
\usepackage[cp1251]{inputenc}

\usepackage[eqsecnum]{cmpj2}
\usepackage{teubner}
\def\qq{\hbox{\fontencoding{LGR}\fontfamily{mtr}\selectfont\foreignlanguage{greek}{\coppa}}}
\def\sqq{{\hbox{\fontencoding{LGR}\fontfamily{mtr}\selectfont\foreignlanguage{greek}{\footnotesize\coppa}}}}

\articletype{Regular article}

\title[Critical Phenomena for Systems under Constraint]%
{Critical Phenomena for Systems under Constraint
}

\author[Izmailian \& Kenna]{Nickolay Izmailian\refaddr{label1,label2} and
        Ralph Kenna\refaddr{label2}}
\addresses{
\addr{label1} Yerevan Physics Institute, Alikhanian Brothers 2, 375036 Yerevan, Armenia
\addr{label2} Applied Mathematics Research Center, Coventry University, Coventry CV1 5FB, England
}

\begin{document}

\maketitle

\begin{abstract}
It is well known that the imposition of a constraint can transform the properties of critical systems. 
Early work on this phenomenon by Essam and Garelick, Fisher, and others, focused on the effects of constraints on the leading critical exponents describing phase transitions. 
Recent work  extended these considerations to critical amplitudes and to exponents governing logarithmic corrections in certain marginal scenarios.
Here these old and new results are gathered and summarised.  
The involutory nature of the transformations between the critical parameters describing ideal and constrained systems are also discussed, paying particular attention to matters relating to universality.

\pacs 64.10.+h, 64.60.-i, 64.60.Bd
\end{abstract}

%%%%%%%%%%%%%%%%%%%%%%%%%%%%%%%%%%%%%%%%%%%%%%%%%%%%%%%%%%%%%%%%%%%%%%%%
%%%%%%%%%%%%%%%%%%%%%%%%%%%%%%%%%%%%%%%%%%%%%%%%%%%%%%%%%%%%%%%%%%%%%%%%
\section{Introduction}
\label{Introduction}
%%%%%%%%%%%%%%%%%%%%%%%%%%%%%%%%%%%%%%%%%%%%%%%%%%%%%%%%%%%%%%%%%%%%%%%%
%%%%%%%%%%%%%%%%%%%%%%%%%%%%%%%%%%%%%%%%%%%%%%%%%%%%%%%%%%%%%%%%%%%%%%%%

The study of thermodynamic systems subject to constraints has a long history.
In 1966, Syozi and Miyazima produced a diluted version of the Ising model and observed that annealed non-magnetic impurities affect the critical behaviour of the model \cite{SM}. 
In particular, the usual infinite critical peak in the specific heat is replaced by a finite cusp.
In 1967, Essam and Garelick quantified the nature of this change as \cite{EG1,EG2}
\begin{equation} 
\alpha_X = -\frac{\alpha}{1-\alpha}.
\end{equation}
Here, $\alpha$ represents the specific heat critical exponent for the ideal (non-diluted) system and $\alpha_X$ is its counterpart for the diluted system. 
If $\beta$ and $\gamma$ similarly represent the magnetisation and susceptibility exponents, Essam and Garelick further showed that these transform to \cite{EG1,EG2}
\begin{equation}
 \beta_X = \frac{\beta}{1-\alpha} , \quad \quad  \gamma_X = \frac{\gamma}{1-\alpha}.
\end{equation}
In 1968, Fisher produced a general theory for critical systems under constraint and the general process linking the ideal critical exponents to those for the constrained system became known as {\emph{Fisher renormalisation}} \cite{Fi68}. 
Because of their continued academic importance and relevance to real systems, phase transitions in constrained systems remained a focus of study \cite{Lushnikov,Aharony,Perk1,Perk2,Perk3}. In recent years the transformation has been extended to deal with other aspects of critical phenomena \cite{us1,us2}.

Because of their experimental accessibility, amplitude terms are important for the description of critical phenomena. 
Unsurprisingly, these also change when a constraint is imposed. 
Perhaps surprisingly, however, the precise nature of this transformation has only recently been studied \cite{us2}. 
Furthermore, in certain marginal circmstances, multiplicative logarithmic corrections also enter into the scaling description at continuous phase transitions. Examples include at the upper critical dimension of spin systems and at the border to regimes where the transition becomes first-order. The exponents of such logarithmic corrections also transform when the system is subjected to a constraint \cite{us2}. 

To give a compact description of all of these various aspects (leading critical exponents, logarithmic corrections and amplitudes), we express the scaling behaviour of the ideal system as follows.
\begin{eqnarray}
 C(t,0) & = &  A_\pm|t|^{-\alpha} |\ln{|t|}|^{\hat{\alpha}} \,,
\label{Ca}
\\
 m(t,0) & = & B |t|^{\beta} |\ln{|t|}|^{\hat{\beta}} \quad {\mbox{for~}} t < 0\,,
\label{mb}
\\
 \chi(t,0) & = & \Gamma_\pm|t|^{-\gamma} |\ln{|t|}|^{\hat{\gamma}} \,, 
\label{cg}
\\
 m(0,h) & = & D h^{\frac{1}{\delta}} |\ln{|t|}|^{\hat{\delta}} \,,
\label{mh}
 \\
 \xi(t,0) & = & N_\pm|t|^{-\nu} |\ln{|t|}|^{\hat{\nu}} \,.
\label{xn}
\end{eqnarray}
Here, $t$ and $h$ refer to the reduced temperature and magnetic field respectively.
The correlation length in the absence of external field is $\xi(t,0)$.
The subscripts $+$ and $-$ refer to amplitudes for $t>0$ and $t<0$, respectively.
In principle we could employ  subscripts for the critical exponents 
and their logarithmic counterparts corresponding to those used for the amplitudes, but 
we suppress these here for simplicity and because the exponents generally coincide on either side of the transition.
Note that Eq.(\ref{Ca}) for the specific heat corresponds to 
an internal energy of leading form
\begin{equation}
 e(t,0) = \pm \frac{A_\pm}{1-\alpha}|t|^{1-\alpha} |\ln{|t|}|^{\hat{\alpha}}.
 \label{ea}
\end{equation}
Finally, and for completeness, we mention that the leading form for the critical correlation function is
\begin{equation}
 G(t=0,h=0;x) = \frac{\Theta}{x^{d-2+\eta}} |\ln{x}|^{\hat{\eta}}
\,.
\label{G0}
\end{equation}

In what follows, we give a comprehensive overview of the effects of the presence of a constraint on the critical exponents (including those of the logarithmic  corrections, when present) and the amplitudes. The critical exponents are universal quantities while the amplitudes are not.
However, certain combinations of amplitudes are universal. 
We show that the renormalisation process (Fisher renormalisation) which transforms the universal critical paramenters is involutary in the sense that applying it twice results in the identity transformation. However, quantities which are not universal do not transform as involutions. We also show that the various scaling relations between the critical parameters (exponents and amplitudes) also hold for the transformed quantities.

In the next section, we summarise the scaling relations for the leading exponents, their logarithmic counterparts and the universal amplitude combinations. 
In Section~3 we apply the renormalisation process and study its effects in Section~4.
We conclude in Section~5.

%%%%%%%%%%%%%%%%%%%%%%%%%%%%%%%%%%%%%%%%%%%%%%%%%%%%%%%%%%%%%%%%%%%%%%%%
%%%%%%%%%%%%%%%%%%%%%%%%%%%%%%%%%%%%%%%%%%%%%%%%%%%%%%%%%%%%%%%%%%%%%%%%
\section{Scaling Relations and Universal Amplitude Combinations}
\label{Second part}
%%%%%%%%%%%%%%%%%%%%%%%%%%%%%%%%%%%%%%%%%%%%%%%%%%%%%%%%%%%%%%%%%%%%%%%%
%%%%%%%%%%%%%%%%%%%%%%%%%%%%%%%%%%%%%%%%%%%%%%%%%%%%%%%%%%%%%%%%%%%%%%%%

The four standard scaling relations  are (see, e.g., Ref.~\cite{Fi98} and references therein)
\begin{eqnarray}
 \alpha + d \nu  & = & 2 \,,
\label{J}
\\
\alpha + 2 \beta + \gamma & = & 2  \,,
\label{R}
\\
 (\delta - 1) \beta & = & \gamma \,,
\label{G}
\\
(2 - \eta) \nu & = & \gamma \,,
\label{F}
\end{eqnarray}
where $d$ represents the dimensionality of the system.
The corresponding scaling relations for the logarithmic-correction exponents are 
\begin{eqnarray}
 \hat{\alpha} + d \hat{\nu} & = & d  \hat{\qq} \,, 
\label{KJJ1}\\
 \hat{\alpha} + \hat{\gamma} & = & 2\hat{\beta}  \,,
\label{KJJ2}\\
  (\delta - 1)  \hat{\beta} + \hat{\gamma} & = & \delta \hat{\delta} \,,
\label{KJJ3}\\
(2-\eta) \hat{\nu} +  \hat{\eta} & = & \hat{\gamma} \,,
\label{KJJ4}
\end{eqnarray}
where $\hat{\alpha}$ is augmented by unity in certain special circumstances
described in Ref.~\cite{us3}. 
The exponent $\hat{\qq}$ (``koppa-hat'') characterises the leading logarithmic correction to the finite-size scaling of the correlation length $\xi_L(0,0) \sim L (\ln{L})^{\hat{\sqq}}$, where $L$ is the finite extent of the system \cite{us4}. 
It is the logarithmic counterpart of the exponent $\qq$, recently introduced to characterise the finite-size correlation length above the upper critical dimension: $\xi_L(0,0) \sim L^{\sqq}$ \cite{us4}.  
%(Since the specific heat is non-divergent above the upper critical dimension where $\qq$ is non-trivial, Fisher renormalisation does not alter the critical exponents there. 
%Therefore  $\qq$ plays 
The relations (\ref{J})--(\ref{F}) for the leading exponents are derived in the appendix where it is also shown that they correspond to the following universal ratios \cite{PHA}:
\begin{eqnarray}
  R_{\xi}  & = & A_\pm N_\pm^d\,,
  \label{Rxi}
  \\
  R_{c}    & = &  \frac{A_\pm \Gamma_\pm}{B^2}\,,
  \label{Rc}
  \\
 R_{\chi}  & = & \frac{ \Gamma_\pm B^{\delta-1}}{D^\delta}\,,
  \label{Rchi}
  \\
  Q        & = &\frac{\Theta N_\pm^{2-\eta}}{\Gamma_\pm}\,,
  \label{Q}
\end{eqnarray}
For the derivation of the logarithmic scaling relations (\ref{KJJ1})--(\ref{KJJ4}), the reader is referred to Refs.~\cite{us3}

In the next section, we examine the effects of constraints on the critical exponents and amplitudes. It will turn out that the renormalised critical exponents obey the same set of scaling relations as their original counterparts and that, when applied to universal quantities, Fisher renormalisation is involutory.

%%%%%%%%%%%%%%%%%%%%%%%%%%%%%%%%%%%%%%%%%%%%%%%%%%%%%%%%%%%%%%%%%%%
%%%%%%%%%%%%%%%%%%%%%%%%%%%%%%%%%%%%%%%%%%%%%%%%%%%%%%%%%%%%%%%%%%%
\noindent
\section{Fisher Renormalisation}
\setcounter{equation}{0}
%%%%%%%%%%%%%%%%%%%%%%%%%%%%%%%%%%%%%%%%%%%%%%%%%%%%%%%%%%%%%%%%%%
%%%%%%%%%%%%%%%%%%%%%%%%%%%%%%%%%%%%%%%%%%%%%%%%%%%%%%%%%%%%%%%%%%%
 
We consider a thermodynamic variable $x$ conjugate to a field $u$, so that
\begin{equation}
 x(t,h,u)
 = \frac{\partial f_X(t,h,u)}{\partial u}
\,.
\label{218}
\end{equation}
Here $f_X (t,h,u)$ represents the  free energy of the  system under constraint and
$u$ represents a quantity such as the chemical potential with $x$ representing the density of annealed non-magnetic impurities. 
The constraint is then expressed in terms of an analytic function as
\begin{equation}
 x(t,h,u)
 = X(t,h,u)
\,.
\label{constraint227}
\end{equation}
One may further assume that the singular part of the free energy of the constrained system is structured analogously to its ideal counterpart $f$, so that
\begin{equation}
 f_X(t,h,u) = f[t^*(t,h,u),h^*(t,h,u)]\,,
\,
\label{start}
\end{equation}
up to a regular background term and in which $t^*$ and $h^*$ are analytic functions \cite{Fi68}. 
The  ideal free energy $f(t,h)$ is recovered if $u$ is fixed at $u=0$.

We assume that 
\begin{equation}
 h^*(t,h,u) = h  {\cal{J}}(t,h,u)\,,
\label{h*}
\end{equation}
so that $h^* = 0 $ when $h=0$.
Then
\begin{equation}
  \frac{\partial  h^*(t,0,u)}{\partial t} = 0 \,, \quad 
  \frac{\partial  h^*(t,0,u)}{\partial u} = 0 \,,
\label{7.2}
\end{equation}
and
\begin{equation}
 \frac{\partial h^*(t,h,u)}{\partial h}
  = 
  {\cal{J}}(t,h,u)
  +
    h
  \frac{\partial  {\cal{J}}(t,h,u)}{\partial h} \,,
\label{7.3}
\end{equation}
so that
\begin{equation}
  \frac{\partial  h^*(t,0,u)}{\partial h} = {\cal{J}}(t,0,u) \,.
\label{7.4}
\end{equation}
For simplicity, we also assume $h \rightarrow -h$ symmetry so that $t^*$ is a function of $h^2$. In that case, 
\begin{equation}
 \frac{\partial t^*(t,h,u)}{\partial h} \propto h\,,
 \label{10.5}
\end{equation}
which vanishes at $h=0$.

%%%%%%%%%%%%%%%%%%%%%%%%%%%%%%%%%%%%%%%%%%%%%%%%%%%%%%%%%%%%%%%%%%%
\subsection{The  critical point}
%%%%%%%%%%%%%%%%%%%%%%%%%%%%%%%%%%%%%%%%%%%%%%%%%%%%%%%%%%%%%%%%%%%

To identify the critical point of the constrained system, one first writes the  magnetization from Eq.(\ref{start}) as
\begin{equation}
 m_X(t,h,u) = \frac{\partial f_X(t,h,u)}{\partial h} 
% & = & \frac{\partial f_X(t,h,u)}{\partial h} 
% = 
%  \frac{\partial f (t^*,h^*)}{\partial h}
%  =
%  \frac{\partial f}{\partial t^*}\frac{\partial t^*}{\partial h}
%                                                                          + % \frac{\partial f}{\partial h^*}\frac{\partial h^*}{\partial h}
% \nonumber
% \\
%              & = &
           = e(t^*,h^*)\frac{\partial t^*}{\partial h}
                                                                         + m(t^*,h^*)  \frac{\partial  h^*(t,0,u)}{\partial h} .
\label{10}
\end{equation}
From Eq.(\ref{10.5}), if the dependency on $h$ is even,
the first term on the right hand side of Eq.(\ref{10}) vanishes at $h=0$.
From Eq.(\ref{7.3}), then
\begin{equation}
 m_X(t,0,u)  =   m [ t^*(t,0,u),0]{\cal{J}}(t,0,u)  \,.
\label{25}
\end{equation}
Now, the critical point of the ideal system is given by the vanishing of $m$.
Assuming that ${\cal{J}}(t,0,u)$ is non-vanishing, Eq.(\ref{25}) gives that
$m_X(t,0,u)  $ vanishes only when $m [ t^*(t,0,u),0] = 0$. 
This means that critical point for the constrained system is given by
\begin{equation}
t^*(t,0,u) = 0 \,.
\end{equation}
(The vanishing of ${\cal{J}}(t,0,u)$ would lead to  two critical points instead of one for the constrained system.)
%This justifies the earlier assumption that the constrained critical point is coincident with the ideal one.
The Taylor expansion for the function ${\cal{J}}(t,h,u)$ about the critical point is 
\begin{equation}
 {\cal{J}}(t,h,u) = J_0 + b_1t + \dots + c_1h + \dots + c_1(u-u_c) + \dots\,,
\label{41}
\end{equation}
where $u_c$ is the critical value of $u$ for the constrained system.
The critical point therefore has  ${\cal{J}}(0,0,u_c) = J_0$.

%%%%%%%%%%%%%%%%%%%%%%%%%%%%%%%%%%%%%%%%%%%%%%%%%%%%%%%%%%%%%%%%%%%
\subsection{The relation between $t^*$ and $t$}
%%%%%%%%%%%%%%%%%%%%%%%%%%%%%%%%%%%%%%%%%%%%%%%%%%%%%%%%%%%%%%%%%%%

The constraint (\ref{constraint227}) determines the relation between $t^*$ and $t$. 
Eq.(\ref{218}) firstly gives
\begin{equation}
 x(t,h,u)  =  \frac{\partial f(t^*,h^*)}{\partial  t^*}\frac{\partial t^*}{\partial u}  +  \frac{\partial f_X(t^*,h^*)}{\partial  h^*}\frac{\partial h^*}{\partial u}.
\label{dinner}
\end{equation}
At $h=0$, the second term on the right vanishes after Eq.(\ref{7.2}). 
Therefore
\begin{equation}
 x(t,0,u) = e(t^*,0) \frac{\partial t^*(t,0,u)}{\partial u}\,.
\label{28}
\end{equation}
This will give a non-trivial relationship between $t^*$ and $t$. 
Expanding  $t^*(t,0,u)$, one has 
\begin{equation}
t^*(t,0,u) = a_1(u-u_c) + \dots\,,
\label{sts}
\end{equation}
where  $u_c$ and the coefficients of the expansion are non-universal.
Therefore
\begin{equation}
 x(t,0,u) =  a_1 e(t^*,0) + \dots
\,,
\label{288}
\end{equation}
which, from Eq.(\ref{ea}), is
\begin{equation}
 x(t,0,u) = \pm a_1 \frac{A_\pm}{1-\alpha} | t^*|^{1-\alpha} |\ln{|t^*|}|^{\hat{\alpha}}+ \dots
\,.
\label{19}
\end{equation}

On the other hand, Taylor expansion of the constraining function gives
\begin{eqnarray}
 X(t,0,u) & = &  X(0,0,u_c) + d_1 (u-u_c) + d_2t + \dots \,,
\label{32}
\\
 & = & X(0,0,u_c) + \frac{d_1}{a_1} t^* + d_2t + \dots \,,
\end{eqnarray}
from (\ref{sts}). 
Comparison with Eq.(\ref{288}) leads to the vanishing of $X(0,0,u_c)$  and
%\footnote{It is satisfying to note that the signs are correct here.}
\begin{equation}
\pm a_1 \frac{A_\pm}{1-\alpha} | t^*|^{1-\alpha} |\ln{|t^*|}|^{\hat{\alpha}}
=
\frac{d_1}{a_1} t^* + d_2t + \dots \,.
\label{22}
\end{equation}
If $\alpha < 0$,  $t^* \sim t$ and the renormalisation is trivial. 
In the case where $\alpha >0$, however, $t$ renormalises to $t^*$ in a non-trivial manner.
To describe this, define 
\begin{equation}
 a = \left[{ \frac{d_2 (1-\alpha)}{a_1}  }\right]^{\frac{1}{1-\alpha}}\,.
\end{equation}
Then the central result is that  the constraint renormalises the  reduced temperature from $t$ to $t^*$, whereby  
\begin{equation}
 |t^*| = a \left({ \frac{|t|}{A_\pm} }\right)^{\frac{1}{1-\alpha}} |\ln{|t|}|^{-\frac{\hat{\alpha}}{1-\alpha}}
\,.
\label{main}
\end{equation}

%%%%%%%%%%%%%%%%%%%%%%%%%%%%%%%%%%%%%%%%%%%%%%%%%%%%%%%%%%%%%%%%%%%
\subsection{Scaling for the constrained system}
%%%%%%%%%%%%%%%%%%%%%%%%%%%%%%%%%%%%%%%%%%%%%%%%%%%%%%%%%%%%%%%%%%%

Eqs.(\ref{start}), (\ref{7.2}) and (\ref{main}) deliver the leading internal energy and specific head for the constrained system as
\begin{equation}
e_X(t,0,u)  =  \frac{\partial f_X(t,0,u)}{\partial t} 
% = \frac{\partial f(t*,0)}{\partial  t^*}\frac{\partial t^*}{\partial t}
% + \frac{\partial f}{\partial  h^*}\frac{\partial h^*}{\partial t}
 =  e(t^*,0) \frac{\partial t^*(t,0,u)}{\partial t}
% + m(t^*,0) \frac{\partial h^*}{\partial t}\,.
 =
 \pm \frac{a^{2-\alpha}}{(1-\alpha)^2} A_\pm^{\frac{-1}{1-\alpha}} |t|^{\frac{1}{1-\alpha}}
 |\ln{|t|}|^{-\frac{\hat{\alpha}}{1-\alpha}}\,,
 \label{eX}
\end{equation}
and
\begin{equation}
  C_X(t,0,u) = \frac{\partial e_X(t,0,u)}{\partial t} = \frac{a^{2-\alpha}}{(1-\alpha)^3}A_\pm^{\frac{-1}{1-\alpha}} |t|^{\frac{\alpha}{1-\alpha}}
  |\ln{|t|}|^{-\frac{\hat{\alpha}}{1-\alpha}}\,,
\end{equation}
respectively
We identify the latter as
\begin{equation}
 C_X(t,0)  =   {A_X}_\pm|t|^{-\alpha_X} |\ln{|t|}|^{\hat{\alpha}_X} \,,
 \label{CX}
\end{equation}
where
\begin{equation}
 \alpha_X = -\frac{\alpha}{1-\alpha}\,,
 \quad \quad \hat{\alpha}_X= -\frac{\hat{\alpha}}{1-\alpha}\,,
 \quad \quad 
  {A_X}_\pm =
  a^{1+\frac{1}{1-\alpha_X}} (1-\alpha_X)^3 A_\pm^{\alpha_X-1}
 \,.
\label{alphaX}
\end{equation}
The last relationship is non-universal since, besides $A_\pm$, $a$ is a non-universal constant.

The magnetization for the constrained system is given by Eqs.(\ref{mb}), (\ref{25}) and (\ref{41}) as  $m_X(t,0,u) = J_0 B |t^*|^{\beta} |\ln{|t^*|}|^{\hat{\beta}}$ for $t<0$. In terms of $t$, we write
\begin{equation}
 m_X(t,0)  =  B_X |t|^{\beta_X} |\ln{|t|}|^{\hat{\beta}_X} \quad {\mbox{for~}} t < 0\,,
 \label{mX}
 \end{equation}
and identify
\begin{equation}
 \beta_X = \frac{\beta}{1-\alpha}\,, \quad \quad 
 \hat{\beta}_X = \hat{\beta} - \frac{\beta \hat{\alpha}}{1-\alpha}\,, \quad \quad 
  B_X = J_0 a^\beta \frac{B}{A_-^{\beta_X}}
 .
\label{betaX}
\end{equation} 

Differentiating Eq.(\ref{10}) with respect to $h$ delivers the susceptibility for the constrained system and, using Eq.(\ref{10.5}) at $h=0$, together with Eqs.~(\ref{7.3}) and (\ref{7.4}), we obtain
$ \chi_X(t,0,u) = J_0^2 \chi (t^*,0) = {\Gamma_X}_\pm |t|^{-\gamma_X}
 |\ln{|t^*|}|^{\hat{\gamma}_X} $, or
\begin{equation}
 \chi_X(t,0)  =  {\Gamma_X}_\pm|t|^{-\gamma_X} |\ln{|t|}|^{\hat{\gamma}_X} \,, 
 \label{chiX}
 \end{equation}
where
\begin{equation}
\gamma_X = \frac{\gamma}{1-\alpha} \,, \quad \quad
\hat{\gamma}_X = \hat{\gamma} + \frac{\gamma \hat{\alpha}}{1-\alpha}\,, \quad \quad
 {\Gamma_X}_\pm =  J_0^2 a^{-\gamma}  A_\pm^{\gamma_X} \Gamma_\pm \,.
\label{ggammaX}
\end{equation}

If $\delta > 1$, the critical isotherm $t=0$ has leading magnetization in field given by Eqs.(\ref{7.3}), (\ref{10.5}) and (\ref{10}) as $m_X(0,h,u) = J_0 Dh^{\frac{1}{\delta}} |\ln{h}|^{\hat{\delta}} $. We identify
\begin{equation}
 m_X(0,h)  =  D_X h^{\delta_X} |\ln{h}|^{\hat{\delta}_X}  \,,
 \label{mhX}
\end{equation}
 with 
\begin{equation}
 \delta_X = \delta\,, \quad \quad 
 \hat{\delta}_X = \hat{\delta}\,, \quad \quad 
  D_X = J_0^{1 + \frac{1}{\delta}} D.
\label{d777}
\end{equation}
The critical exponents are therefore unchanged but the amplitude undergoes a transformation.

The correlation length  renormalises in a similar way to the susceptibility since
$\xi_X(t) = \xi (t^*) = N_\pm |t^*|^{-\nu}|\ln{|t^*|}|^{-\hat{\nu}}$.
We write
\begin{equation}
 \xi_X(t,0)  
              = {N_X}_\pm |t|^{-\nu_X}|\ln{|t|}|^{-\hat{\nu}_X}\,,
\label{xiX}
\end{equation}
where
\begin{equation}
\nu_X = \frac{\nu}{1-\alpha} \,, \quad \quad
\hat{\nu}_X = \hat{\nu} + \frac{\nu \hat{\alpha}}{1-\alpha}\,, \quad \quad
 {N_X}_\pm =  a^{-\nu} A_{\pm}^{\nu_X} N_\pm .
\label{nuX}
\end{equation}

Finally, the correlation function is obtainable by differentiating the free energy with respect to two local fields $h_1=h(x_1)$ and $h_2=h(x_2)$. One obtains
\[
 G_X(t,h,u;x) = \frac{\partial^2 f_X(t,h,u)}{\partial h_1 \partial h_2} 
              = J_0^2 \frac{\partial^2 f(t^*,h^*)}{\partial h_1^* \partial h_2^*} 
               = J_0^2  G(t*,h^*,x).
\]
Setting $t^*=t=h^*=h=0$, delivers $G_X(0,0,u;x) = J_0^2 G(0,0,x)$. Writing
\begin{equation}
 G_X(0,0,x)  =  \frac{\Theta_X}{x^{d-2+\eta_X}} |\ln{x}|^{\hat{\eta}_X} \,,
 \label{G00X}
\end{equation}
we identify
\begin{equation}
\eta_X = \eta \,, \quad \quad
\hat{\eta}_X = \hat{\eta} \,, \quad \quad
 \Theta_X = J_0^2 \Theta
 .
 \label{ThetaXXX}
\end{equation}

We have observed that neither the in-field magnetisation nor the correlation function exhibit non-trivial renormalisation of the critical exponents.
The former is the case by construction and the latter is so because it is defined at the critical point. Likewise, the exponents $\qq$ and $\hat{\qq}$ governing finite-size scaling of the correlaton length do not change under Fisher renormalisation, so that $\qq_X=\qq$ and $\hat{\qq}_x=\hat{\qq}$.

%%%%%%%%%%%%%%%%%%%%%%%%%%%%%%%%%%%%%%%%%%%%%%%%%%%%%%%%%%%%%%%%%%%
%%%%%%%%%%%%%%%%%%%%%%%%%%%%%%%%%%%%%%%%%%%%%%%%%%%%%%%%%%%%%%%%%%%
\section{Properties of Renormalised Scaling Parameters}
\label{next part}
%%%%%%%%%%%%%%%%%%%%%%%%%%%%%%%%%%%%%%%%%%%%%%%%%%%%%%%%%%%%%%%%%%%
%%%%%%%%%%%%%%%%%%%%%%%%%%%%%%%%%%%%%%%%%%%%%%%%%%%%%%%%%%%%%%%%%%%

It is straightforward to verify that if the critical exponents for the ideal system satisfy the scaling relations (\ref{J})--(\ref{F}), the renormalised exponents for the constrained system do likewise.
(This observation for the Essam-Fisher relation (\ref{R}) was already made in Ref.~\cite{EG1}.)  
The same statement applies to the scaling relations for logarithmic corrections (\ref{KJJ1})--(\ref{KJJ4}).

Fisher renormalisation applied to the universal critical exponents is involutory. 
This  means that renormalisation of renormalised exponents delivers the pure values.
For example, $\gamma_{XX} = \gamma_X/(1-\alpha_X) = \gamma$ and 
$\hat{\gamma}_{XX} = \hat{\gamma}_X + \gamma_X \hat{\alpha}_X/(1-\alpha_X) = \hat{\gamma}$.
However, the same starement does not apply to the amplitudes. 
For example, two successive applications of Ew.~(\ref{ggammaX}) give ${\Gamma_{XX}}_\pm$  different  from $\Gamma_\pm$.

Of course, the critical exponents, for which the transformation is involutory, are universal, whereas the critical amplitudes are not.
This observation prompts one to investigate the nature of the universal combinations 
(\ref{Rxi})--(\ref{Q}) under Fisher renormalisation.
The non-universal terms $J_0$ and $a$, which characterise  the transformations of the individual amplitude terms,  drop out of the transformations of the universal combinations through the scaling relations (\ref{J})--(\ref{F}).
%These also transform non-trivially uner Fisher renormalization. 
The universal amplitude combinations  transform as 
\begin{eqnarray}
 {R_X}_c & = & \frac{1}{(1-\alpha)^3}R_c\,, 
 \label{Rctrans} \\
 {R_X}_\chi & = & R_\chi \,,
 \label{Rchitrans} \\
 {R_X}_\xi & = & \frac{1}{(1-\alpha)^3} R_\xi \,,
 \label{Rxitrans} \\
 Q_X & = & Q \,,
 \label{RQtrans} \\
 Z_X & = & \frac{Z}{U_0^{\Delta_X}} \,.
 \label{RZtrans} \\
\end{eqnarray}
Two successive applications of these transformations confirm the involutory nature of these universal combinations.

%%%%%%%%%%%%%%%%%%%%%%%%%%%%%%%%%%%%%%%%%%%%%%%%%%%%%%%%%%%%%%%%%%%
%%%%%%%%%%%%%%%%%%%%%%%%%%%%%%%%%%%%%%%%%%%%%%%%%%%%%%%%%%%%%%%%%%%
\section{Conclusions}
\label{last part}
%%%%%%%%%%%%%%%%%%%%%%%%%%%%%%%%%%%%%%%%%%%%%%%%%%%%%%%%%%%%%%%%%%%
%%%%%%%%%%%%%%%%%%%%%%%%%%%%%%%%%%%%%%%%%%%%%%%%%%%%%%%%%%%%%%%%%%%

Fisher renormalization, which generalises an earlier theory of Essam and Garelick is a staple of the established theory of critical phenomena.
The early work by these authors was extended in recent years to encompass critical amplitudes and the exponents which govern logarithmic corrections to scaling, when present.
Here, a comprehensive treatment of all of these various elements has been given.
we also observe that the involutory nature of the renormalisation process is intrinsically  linked to universality.

%%%%%%%%%%%%%%%%%%%%%%%%%%%%%%%%%%%%%%%%%%%%%%%%%%%%%%%%%%%%%%%%%%%
%%%%%%%%%%%%%%%%%%%%%%%%%%%%%%%%%%%%%%%%%%%%%%%%%%%%%%%%%%%%%%%%%%%
\section{Acknowledgments}
\label{Acknowledgments}
%%%%%%%%%%%%%%%%%%%%%%%%%%%%%%%%%%%%%%%%%%%%%%%%%%%%%%%%%%%%%%%%%%%
%%%%%%%%%%%%%%%%%%%%%%%%%%%%%%%%%%%%%%%%%%%%%%%%%%%%%%%%%%%%%%%%%%%
The work was supported by a Marie Curie IIF (Project no. 300206-RAVEN)
and IRSES (Projects no. 295302-SPIDER and 612707-DIONICOS) within 7th European Community Framework
Programme and by the grant of the Science Committee of the Ministry of Science and
Education of the Republic of Armenia under contract 13-1C080.

%%%%%%%%%%%%%%%%%%%%%%%%%%%%%%%%%%%%%%%%%%%%%%%%%%%%%%%%%%%%%%%%%%
\appendix
\section{Appendix: Universal amplitude Combinations}
\label{Appendix}
\setcounter{equation}{0}
%%%%%%%%%%%%%%%%%%%%%%%%%%%%%%%%%%%%%%%%%%%%%%%%%%%%%%%%%%%%%%%%%%

To identify the universal amplitude combinations, we begin with the standard scaling form for the  free energy and correlation length \cite{PHA,Fi98}
\begin{eqnarray}
 f(t,h)   & = &  b^{-d} Y(K_t b^{y_t} t, K_h b^{y_h}h)\,, 
 \label{A1} \\
 \xi(t,h) & = &  b      X(K_t b^{y_t} t, K_h b^{y_h}h)\,.
\label{A2}
\end{eqnarray}
The scaling functions $Y$ and $X$ are universal and all the non-universality is contained in the metric factors  $K_t$ and $K_h$.

Differentiating Eq.(\ref{A1}) with respect to $h$ delivers the scaling form for the magnetization as
\begin{equation}
 m(t,h) = b^{-d+y_h} K_h Y^{(h)}(K_t b^{y_t} t, K_h b^{y_h} h )\,,
\label{A3}
\end{equation}
where the parenthesized superscript signifies appropriate differentiaton of the scaling function.
Setting $h=0$ and chosing 
\begin{equation}
 b = K_t^{-\frac{1}{y_t}} |t|^{-\frac{1}{y_t}}\,
 \label{A4}
\end{equation}
gives the spontaneous magnetization 
$ m(t,0) = B(-t)^\beta$, for $t<0$, in which
\begin{equation}
 \beta = \frac{d-y_h}{y_t}
 \quad
 {\mbox{and}}
 \quad 
  B = K_t^\beta K_h Y^{(h)}(1,0)\,.
\label{A7}
\end{equation}
On the other hand, setting  $t=0$ in Eq.(\ref{A3}) and choosing 
\begin{equation}
 b = K_h^{-\frac{1}{y_h}} h^{-\frac{1}{y_h}},
 \end{equation}
we obtain 
$ m(0,h) = Dh^{{1}/{\delta}}$
in which
\begin{equation}
  \frac{1}{\delta} = \frac{d-y_h}{y_h}
  \quad
  \mbox{and}
  \quad
   D = K_h^{1 + \frac{1}{\delta}}Y^{(h)}(0,1)
\,.
\label{A12}
\end{equation}
The susceptibility is obtained by  differentiating Eq.(\ref{A3}) with respect to $h$. Again setting $h=0$ and  using Eq.(\ref{A4}), one finds
$ \chi(t,0) = \Gamma_\pm |t|^{-\gamma}$ where
\begin{equation}
\gamma = \frac{2y_h-d}{y_t} 
\quad
{\mbox{and}}
\quad
\Gamma_\pm =  K_t^{-\gamma} K_h^{2} Y^{(hh)}(\pm 1 , 0)\,. 
\label{A18}
\end{equation}
For the specific heat, differentiate (\ref{A1}) twice with respect to $t$ and again use Eq.(\ref{A4}) to find $ C(t,0) = A_\pm |t|^{-\alpha}$ with
\begin{equation}
 \alpha  = 2 - \frac{d}{y_t}
 \quad
 {\mbox{and}}
 \quad
  A_{\pm}  = K_t^{2-\alpha} Y^{(tt)}(\pm 1,0) \,.
\label{A23}
\end{equation}

From Eqs.(\ref{A7}) and (\ref{A12}), we can express $y_t$ and $y_h$ in terms of $\beta$ and $\delta$,
\begin{equation}
 y_t  =  \frac{d}{\beta}\frac{1}{\delta+1}
 \quad
 {\mbox{and}}
 \quad
  y_h  =  \frac{d\delta }{\delta + 1}\,.
 \label{A26}
\end{equation}
Similarly, using Eqs.(\ref{A7}) and (\ref{A12}) we can express $K_t$ and $K_h$ in terms of $B$ and $D$,
\begin{equation}
 K_t =  \left[{\frac{B}{Y^{(h)}(1,0)}}\right]^{\frac{1}{\beta}} 
 \left[{\frac{D}{Y^{(h)}(0,1)}}\right]^{-\frac{1}{\beta}\frac{\delta}{1+\delta}}  
 \quad
 \mbox{and}\quad
 K_h  =  \left[{\frac{D}{Y^{(h)}(0,1)}}\right]^{\frac{\delta}{\delta + 1}}  \,.
 \label{A27}
 \end{equation}
Here, the $Y^{(h)}$ are universal and the amplitudes $B$ and $D$ are not.

Finally, expressing $\alpha$ and $\gamma$ in terms of $\beta$ and $\delta$ through Eqs.(\ref{A18}) and (\ref{A23}) delivers the static scaling relations (\ref{R}) and (\ref{G}). 
Correspondingly, one can express $A_\pm$ and $\Gamma_\pm$ in terms of $B$ and $D$,
\begin{eqnarray}
 \Gamma_\pm & = & 
 \frac{Y^{(hh)}(\pm 1,0)}{
                        \left[{Y^{(h)}(1,0)}\right]^{\frac{1}{\beta}} 
                        \left[{Y^{(h)}(0,1)}\right]^{\delta} 
                        }
                        B^{1-\delta}
                        D^{\delta} \,,
 \label{A31}
 \\
 A_\pm & = & 
 \left[{Y^{(h)}(1,0)}\right]^{-(\delta + 1)}
 \left[{Y^{(h)}(0,1)}\right]^{\delta}
 Y^{(tt)}(\pm 1, 0)
 \frac{B^{\delta+1}}{D^\delta}
  \,.
 \label{32}
 \end{eqnarray}
From the first of these, $\Gamma_\pm B^{\delta-1}/D^\delta$ is a universal combination of universal factors. This is $R_\chi$ in Eq.(\ref{Rchi}).
From the second, the ratio $A_\pm D^\delta/B^{\delta+1}$ is universal. 
Or, combining with Eq.(\ref{A12}), the quantity $R_c$ in Eq.(\ref{Rc}) is seen to be universal.

From Eqs.(\ref{A2}) and (\ref{A4}), 
the correlation length is
$ \xi(t,0) = N_{\pm} |t|^{-\nu}$ where
\begin{equation}
 \nu = \frac{1}{y_t}
 \quad
 \mbox{and}
 \quad
  N_\pm = K_t^{-\frac{1}{y_t}}X(\pm 1,0).
 \label{A35}
\end{equation}
From Eq.(\ref{A23}) the first of these is the hyperscaling relation (\ref{J}).
To connect $N_\pm$ to the other amplitudes, one can exploit the relatonship between the susceptibility and the correlation function,
\begin{equation}
 \chi = \int_0^{\xi}{G(x) x^{d-1} dx} = \Theta \xi^{2-\eta},
\end{equation}
from which Fisher's scaling relation ({\ref{F}) follows, along with
\begin{equation}
 \Gamma_\pm = \Theta N_\pm^{2-\eta}.
 \label{A42}
\end{equation}
The combination $Q=\Theta N_{\pm}^{2-\eta}/\Gamma_{\pm}$ of Eq.(\ref{Q}) is therefore universal.
Similarly, the universality of $R_\xi$ in Eq.(\ref{Rxi}) can be explained through the hyperscaling relation 
$f(t,0) = A_\pm |t|^{2-\alpha}/(2-\alpha)(1-\alpha) \sim \xi^d(t,0)  = (N_\pm|t|^{-\nu})^d$.

%\end{document}

%% Type in your references using {thebibliography} environment
%% or create them from your bibtex database using cmpj.bst style (experimental).

%\bibliographystyle{cmpj}
%\bibliography{mybibdb}

\begin{thebibliography}{99}


%--------------------------------------------------------------------------
\bibitem{SM}
I.~Syozi and S.~Miyazima, Prog. Theor. Phys. {\bf{36}} (1966) 1083.
%-1094.
% A Statistical Model for the Dilute Ferromagnet


%--------------------------------------------------------------------------
\bibitem{EG1}
J.W.~Essam and H.~Garelick, Proc. Phys. Soc. {\bf{92}} (1967) 136.
%Critical behaviour of a soluble model of dilute ferromagnetism



%--------------------------------------------------------------------------
\bibitem{EG2}
H.~Garelick and J.W.~Essam, J. Phys. C (Proc. Phys. soc.) Ser.~2, Vol.~1 (1968) 1588.
%-1595
%Critical behaviour of the three-dimensional Ising model specific heat below Tc

%--------------------------------------------------------------------------


\bibitem{Fi68}
M.E.~Fisher, Phys. Rev. {\bf{176}} (1968) 257.
%-XXX.


%--------------------------------------------------------------------------
\bibitem{Lushnikov}
A.A.~Lushnikov, Phys. Lett. A {\bf{27}} (1968) 158; Sov. Phys. JETP 29 (1969) 120.


%--------------------------------------------------------------------------
\bibitem{Aharony}
A.~Aharony, J. Magn. Magn. Mater. {\bf{7}} (1978) 215.


%--------------------------------------------------------------------------

\bibitem{Perk1}
  H.W. Capel, J.H.H. Perk, and L.W.J. den Ouden,
%  Critical-exponent renormalization and first-order transitions,
  Phys. Lett. A {\bf{66}} (1978) 437.
%-439.

%--------------------------------------------------------------------------

\bibitem{Perk2}
  H.W. Capel, L.W.J. den Ouden, and J.H.H. Perk,
%  Stability of critical behaviour, critical-exponent renormalization and first-order transitions,
  Physica A {\bf{95}} (1979) 371.
%-416.

%--------------------------------------------------------------------------

\bibitem{Perk3}
  L.W.J. den Ouden, H.W. Capel, and J.H.H. Perk,
%  Critical-exponent renormalization, first-order transitions,  and demagnetizing effects for Schofield's linear model,
  Physica A {\bf{105}} (1981) 53.
%-85.



%--------------------------------------------------------------------------

\bibitem{us1}
R. Kenna, H.-P Hsu and C. von Ferber, 
%Fisher Renormalization for Logarithmic Corrections, 
J. Stat. Mech. (2008) L10002.


%--------------------------------------------------------------------------

\bibitem{us2}
N.Sh. Izmailian and R. Kenna, arXiv/1402.4673 (to appear in JSTAT, 2014).


%--------------------------------------------------------------------------

\bibitem{Fi98}
M.E.~Fisher, Rev. Mod. Phys. {\bf{70}} (1998) 653.



%--------------------------------------------------------------------------
\bibitem{us3}
R. Kenna, 
%Universal scaling relations for logarithmic-correction exponents, 
in "Order, Disorder, and Criticality: Advanced Problems of Phase Transition Theory", 
Yu. Holovatch (editor). vol. 3 World Scientific, Singapore, 2012.


%--------------------------------------------------------------------------
\bibitem{us4}
R. Kenna and B. Berche, 
%A new critical exponent q and its logarithmic counterpart, 
Condensed Matter Physics {\bf{16}} (2013) 23601.
%: 1-12.


%--------------------------------------------------------------------------


\bibitem{PHA}
V.~Privman, P.C.~Hohenberg and A.~Aharony, 
%{\emph{Universal Critical-Point Amplitude Ratios}\/}, 
in {\emph{Phase Transitions and Critical Phenomena}} Vol.{\bf{14}}
(Academic, New York, 1991), 
eds. C.~Domb and J.L.~Lebowitz,~pp~1-134.



%--------------------------------------------------------------------------


\end{thebibliography}

%
%% If you have problems with typesetting in ukrainian uncomment lines below.
%
  \lastpage
  \end{document}